\pdfoutput=1

\documentclass[11pt]{article}

\usepackage[]{acl}

\usepackage{times}
\usepackage{latexsym}

\usepackage[T1]{fontenc}

\usepackage[utf8]{inputenc}

\usepackage{microtype}

\usepackage{inconsolata}
\usepackage{subcaption} 
\usepackage{tikz}
\usepackage{amsmath}
\usepackage{circledsteps}
\usepackage{booktabs}
\usepackage{tabularx}
\usepackage{threeparttable}
\usepackage{multirow}
\usepackage{makecell}
\usepackage{float}
\usepackage{amssymb}
\usepackage{enumitem}
\usepackage{xcolor} 
\usepackage{colortbl} 
\usepackage{graphicx}
\usepackage{booktabs}
\usepackage{array}
\usepackage{changepage}
\usepackage{stfloats}

\definecolor{customred}{HTML}{FF5314}
\definecolor{customgreen}{HTML}{7CBB00}

\newcommand{\ModName}{IRBinDiff}

\title{Binary Code Similarity Detection via Graph Contrastive Learning on Intermediate Representations}

\author{
 \textbf{Xiuwei Shang\textsuperscript{1}},
 \textbf{Li Hu\textsuperscript{1}},
 \textbf{Shaoyin Cheng\textsuperscript{1,2,*}},
 \textbf{Guoqiang Chen\textsuperscript{3}},
 \textbf{Benlong Wu\textsuperscript{1}},
\\
 \textbf{Weiming Zhang\textsuperscript{1,2}},
 \textbf{Nenghai Yu\textsuperscript{1,2}}
\\
 \textsuperscript{1}University of Science and Technology of China \\
 \textsuperscript{2}Anhui Province Key Laboratory of Digital Security 
 \textsuperscript{3}QI-ANXIN Technology Research Institute
\\
 \texttt{\{shangxw,pdxbshx,dizzylong\}@mail.ustc.edu.cn} \\
 \texttt{\{sycheng,zhangwm,ynh\}@ustc.edu.cn}  \
 \texttt{guoqiangchen@qianxin.com} 
}

\begin{document}
\maketitle
\begin{abstract}

Binary Code Similarity Detection (BCSD) plays a crucial role in numerous fields, including vulnerability detection, malware analysis, and code reuse identification. 
As IoT devices proliferate and rapidly evolve, their highly heterogeneous hardware architectures and complex compilation settings, coupled with the demand for large-scale function retrieval in practical applications, put forward higher requirements for BCSD methods. 
In this paper, we propose \ModName, which mitigates compilation differences by leveraging LLVM-IR with higher-level semantic abstraction, and integrates a pre-trained language model with a graph neural network to capture both semantic and structural information from different perspectives. By introducing momentum contrastive learning, it effectively enhances retrieval capabilities in large-scale candidate function sets, distinguishing between subtle function similarities and differences.
Our extensive experiments, conducted under varied compilation settings, demonstrate that \ModName~outperforms other leading BCSD methods in both One-to-one comparison and One-to-many search scenarios.

\end{abstract}

\section{Introduction}

Binary code similarity detection (BCSD) aims to determine whether two binary code snippets are similar in functional semantics, which holds significant importance in various domains such as vulnerability detection \cite{david2018firmup}, malware analysis \cite{nguyen2023binary}, and software supply chain analysis \cite{basit2005detecting}, etc. Particularly, with the explosive growth of IoT devices \cite{Statista}, the high heterogeneity of hardware architectures and software platforms across different devices has led to an increase in the diversity and complexity of binary files, thus posing higher demands on BCSD techniques.

In recent years, with the advancements in natural language processing technologies, deep learning-driven BCSD methods have gradually emerged \cite{du2023review, haq2021survey}. These typically embed binary code in assembly form into high-dimensional vector spaces and evaluate the functional similarity of code snippets by calculating vector similarities. For instance, VulSeeker \cite{gao2018vulseeker}, Gemini \cite{xu2017neural}, and OrderMatters leverage graph embedding techniques to encode the control flow graph of binary code into vectors, capturing control flow or data dependencies, while Asm2Vec \cite{ding2019asm2vec}, SAFE \cite{massarelli2019safe}, and PalmTree \cite{li2021palmtree} employ language models to directly learn representations from instruction sequences, capturing instruction order and contextual semantics.

\begin{figure*}[t]
  \centering
  \includegraphics[width=0.9\linewidth]{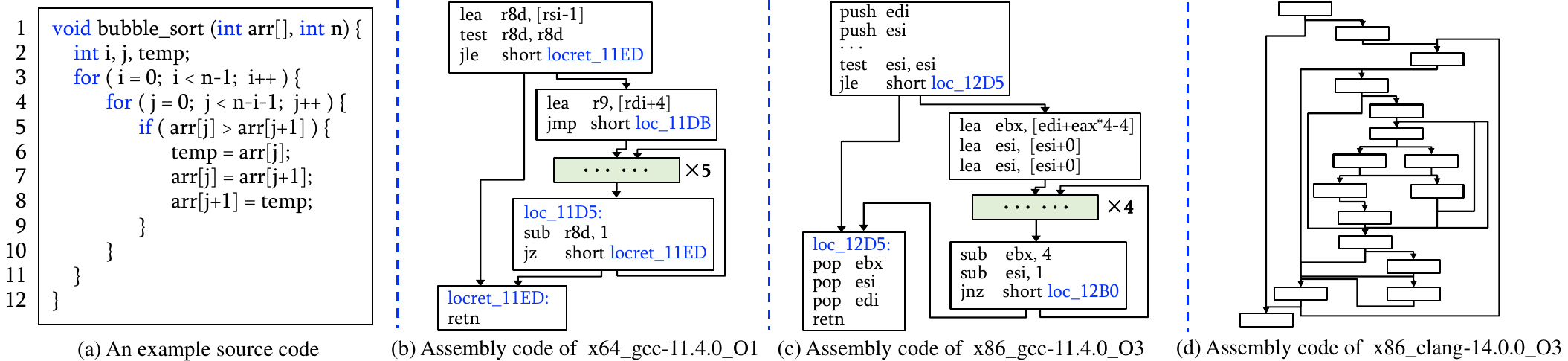}
  \vspace{-1.0ex}
  \caption{Code Example. (a) is the source code for bubble sort, and (b) (c) (d) are the assembly code with control-flow of the binary compiled with x64\_gcc-11.4.0\_O1, x86\_gcc-11.4.0\_O3, and x86\_clang-14.0.0\_O3, respectively.} 
  \vspace{-2ex}
  \label{fig:background}
\end{figure*}

Although these methods have shown significant potential in capturing syntactic and semantic features of binary code, they still face challenges in complex and diverse practical application scenarios:
(1) \textbf{Complex compilation options}: IoT device firmware is typically based on different architectures and uses various compilers and optimization options, resulting in binaries that may appear entirely different in form, even when compiled from the same codebase, as shown in the example in Figure \ref{fig:background}. Therefore, BCSD methods must be robust to such complex compilation settings.
(2) \textbf{Large-scale candidate function retrieval}: In real-world large-scale firmware analysis, it is often necessary to retrieve a few similar functions from a vast number of unrelated functions. This requires BCSD methods to have the ability to capture subtle semantic differences in large-scale datasets.

In this paper, we present \ModName, a novel intermediate representation-based graph contrastive learning method, which aims to address the aforementioned challenges and support real-world binary code similarity detection. 
Firstly, by lifting binary files to LLVM-IR, a flexible and architecture-independent intermediate representation, we effectively mitigate the discrepancies caused by underlying hardware and compilation options \cite{barchi2019code}. LLVM-IR provides a more unified and high-level semantic abstraction of machine code, making it an ideal choice for handling complex compilation settings, thereby enabling the model to apply to highly heterogeneous scenarios such as IoT device firmware. 
Secondly, \ModName~combines a pre-trained language model and a graph neural network to extract functional features from both semantic and structural aspects. The pre-trained language model captures the semantic and syntactic patterns in the serialized representation of simplified and normalized LLVM-IR instructions, while the graph neural network encodes the extracted control-flow graph, capturing the structural information and dependencies within the function.
Finally, we introduce momentum contrastive learning to enhance the model's retrieval capabilities in large-scale candidate functions. By maintaining a dynamic queue of a large number of negative samples, the model can learn from a more diverse set of negatives, effectively distinguishing between semantically similar functions and unrelated ones. Additionally, the momentum update mechanism gradually adjusts the parameters of key encoder, ensuring stability throughout the training process.

Our contributions can be summarized as follows:
\begin{itemize}[left=0.2cm] 
    \setlength{\itemsep}{0pt}  
    \vspace{-0.5ex}
    \item \textbf{Comprehensive Pipeline}. We construct a comprehensive pipeline that lifts binary code to LLVM-IR to address the challenges posed by complex compilation options, while extracting and optimizing its syntactic structure and semantic information. 
    \vspace{-0.5ex}
    \item \textbf{Innovative Methodology}. We introduce \ModName, which combines a pre-trained language model and a graph neural network to deeply learn the functional features of LLVM-IR from the semantic and structural levels, respectively, and utilizes momentum contrastive learning to enhance its ability to capture subtle differences in a large set of candidate functions.
    \item \textbf{Effective Experimental}. We conduct extensive experiments across various compilation options, demonstrating that \ModName~outperforms baselines in both One-to-one comparison and One-to-many search scenarios.
\end{itemize}

\vspace{-1.5ex}
\section{Related Works} \label{sec:relatedworks}
\vspace{-0.8ex}
\subsection{Binary Code Similarity Detection (BCSD)}
\vspace{-0.5ex}

Traditional BCSD methods often rely on raw bytecode or manually crafted rules to assess code similarity. For example, $\alpha$diff \cite{liu2018alphadiff} uses raw binary bytes as input in a siamese network based on convolutional neural networks to compute similarity. However, it fails to capture the program's unique logic and functional features. Methods like discovRE \cite{eschweiler2016discovre} and BinFinder \cite{qasem2023binary} extract statistical features of function, such as constants and strings, for similarity detection, while Gemini \cite{xu2017neural} and VulSeeker \cite{gao2018vulseeker} use hand-crafted features to represent basic blocks and apply graph embedding techniques for function-level similarity. However, these feature-based approaches overly rely on domain expertise and struggle to generalize to diverse applications

Recently, inspired by natural language processing techniques, representation learning-based BCSD methods have gradually become mainstream. INNEREYE \cite{zuo2018neural} and SAFE \cite{massarelli2019safe} treat an assembly instruction as a "word" and use the Word2Vec \cite{mikolov2013efficient} model for representation, while Asm2Vec \cite{ding2019asm2vec} and PalmTree \cite{li2021palmtree} take a finer-grained approach by treating instructions as "sentences" and opcodes or operands as "words," leveraging sequence language models. In this paper, we adopt a representation learning approach, focusing on both instruction sequences and control-flow structures.

\begin{figure*}[htb]
  \centering
  \includegraphics[width=0.88\linewidth]{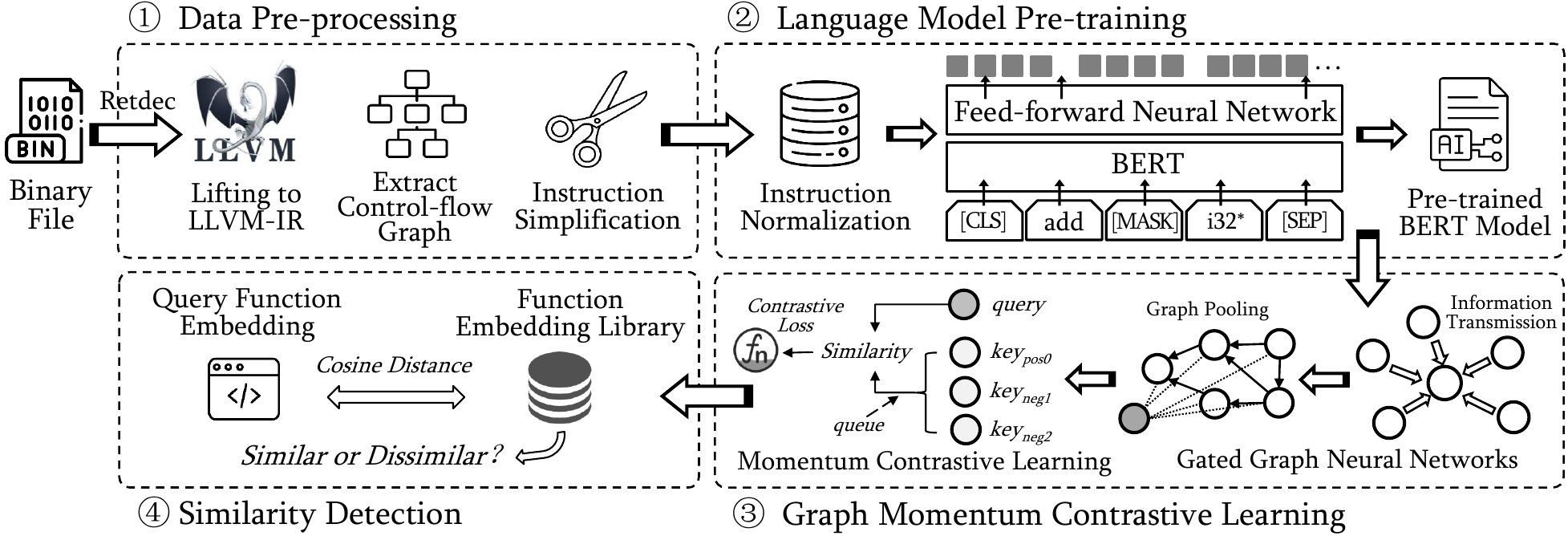}
  \vspace{-0.5ex}
  \caption{Overview framework of \ModName.} 
  \vspace{-2.2ex}
  \label{fig:overview}
\end{figure*}

\vspace{-0.8ex}
\subsection{LLVM-IR}
\vspace{-0.3ex}

LLVM-IR is a typical, well-formatted universal intermediate representation. During the compilation process, it bridges source code and machine code, and can provide a common optimization platform for multiple high-level languages and compiler backends \cite{lattner2004llvm}.

Compared to directly handling assembly code, LLVM-IR offers a higher level of semantic abstraction, allowing binaries from different architectures and compilation settings to be lifted to a unified representation. Additionally, its good readability and standardized syntax enable models to more accurately extract core semantics. The specific example of LLVM-IR is shown in Appendix \ref{sec:cfgappendix}.

\vspace{-0.6ex}
\section{\ModName}
\vspace{-1.2ex}
This section elucidates the rationale and architecture of our designed method \ModName, which consists of four parts, as illustrated in Figure \ref{fig:overview}.

\vspace{-0.6ex}
\subsection{Data Pre-processing} \label{sec:datapre}

\noindent\textbf{Lifting Binary to LLVM-IR.}
For binary files, we use the reverse engineering tool Retdec \cite{kvroustek2017retdec} for decompilation, which calls bin2llvmir to lift the binary to the LLVM-IR format, obtaining files with the \texttt{.ll} extension.

\noindent\textbf{Extract Control-flow Graph.}  
Previous works \cite{xu2017neural, gao2018vulseeker, ding2019asm2vec} have shown that structural information in binary code, such as control-flow graph (CFG), provides crucial semantic insights for similarity detection. To this end, we employ regular expression matching to parse LLVM-IR files, performing function boundary division, basic block division, and CFG extraction. The detailed process is explained in Appendix \ref{sec:cfgappendix}. By doing so, we can represent each function in LLVM-IR as a graph, where the nodes represent basic blocks and the edges represent control flow relationships between these blocks.

\noindent\textbf{Instruction Simplification.}  
The instruction sequences in LLVM-IR contain many instructions unrelated to function semantics, which not only introduce additional computational overhead but also dilute the importance of key instructions, thereby impairing embedding performance. To address this, we empirically devise an instruction simplification strategy, pruning redundant instructions while preserving essential ones. This approach, detailed in Appendix \ref{sec:simplifyappendix}, enables the language model to extract more precise semantic information from functions.

\vspace{-0.6ex}
\subsection{Language Model Pre-training} \label{sec:languagemodel}

\noindent\textbf{Instruction Tokenization and Normalization.} 
IR instructions have relatively complex internal structures, and fully understanding these internal details is crucial for representing the instructions. Therefore, instead of treating an entire instruction as a single word, we apply a fine-grained tokenization strategy, regarding each instruction as a sentence and each token as a word. Specifically, we split the instructions based on spaces, punctuation (e.g., "\texttt{,}" "\texttt{()}", "\texttt{[]}", "\texttt{\{\}}"), and underscores "\texttt{\_}". For instance, give a LLVM-IR instruction "\texttt{\%16 = call i32 @\_cxa\_begin\_catch (i32* \%result)}", we divide it into "\texttt{\%16}", "\texttt{=}", "\texttt{call}", "\texttt{i32}", "\texttt{@}", "\texttt{cxa}", "\texttt{begin}", "\texttt{catch}", "\texttt{(}", "\texttt{i32*}", "\texttt{\%result}", "\texttt{)}".

To mitigate the data sparsity and out-of-vocabulary (OOV) issues caused by the diversity of identifiers and constants, inspired by related research \cite{gao2021lightweight}, we empirically establish instruction normalization rules: (1) Replace all identifiers of the form "\%dec\_label\_pc\_xxxx" with \texttt{<label>}. (2) Replace global variables, such as "@global\_var\_73008" with \texttt{<global>}. (3) Treat numeric strings with absolute values less than 1024 as constant values and replace them with \texttt{<Positive>} and \texttt{<Negative>} according to their signs. (4) Treat numeric strings with absolute values greater than 1024 as addresses and replace them with \texttt{<Address>}. (5) Remove the prefix and suffix numbers from identifiers like "reg2mem", "reload", "\%storemerge", "\%brmerge", "select", and "thread". An intuitive explanation is given in Appendix \ref{sec:pretrainappendix}.

\noindent\textbf{Random Walk Sampling.}  
To generate training inputs for the language model, we use a random walk \cite{li2015random} algorithm to sample instruction pairs, which input is a graph structure. In Section \ref{sec:datapre}, we have already extracted the CFG of the function, where each node represents a basic block, and the instructions within a basic block are executed sequentially. We can further extend this to a CFG where each node represents a single instruction. We set each instruction as the starting point for sampling, and randomly select a node $w$ from the successors of the current node $v$ as the sampling endpoint. The probability of a random walk from node $v$ to its successor node $w$ is given by:
\begin{equation}
\footnotesize
P^{RW}_{v,w} = 
\begin{cases} 
\frac{1}{d(v)}, & \text{if } w \text{ after } v \\ 
0, & \text{otherwise} 
\end{cases}
\end{equation}
where $d(v)$ denotes the number of successors of node $v$. After performing random walk sampling, we obtain a corpus of instruction pairs.

\begin{figure}[t]
  \centering
  \includegraphics[width=0.95\linewidth]{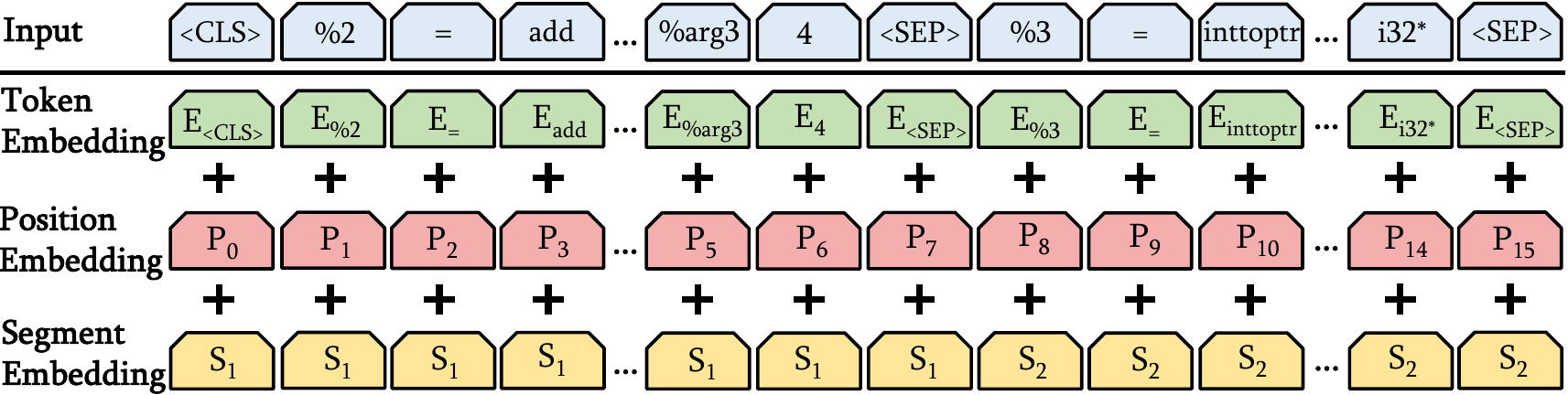}
  \vspace{-0.5ex}
  \caption{Input representation of language model.} 
  \vspace{-1.8ex}
  \label{fig:representation}
\end{figure}

\noindent\textbf{Language Model Training.}  
Our LLVM-IR language model is based on BERT \cite{devlin2018bert}, which is a multi-layer bidirectional transformer \cite{vaswani2017attention} encoder that takes instruction pairs as input. As shown in Figure \ref{fig:representation}, the first token is [CLS], which marks the begining of the sequence, and token [SEP] is used to separate the instruction pairs. Additionally, we combine position embedding, segment embedding with token embedding as the input of the language model. The position embedding represent the position of token within the input sequence, while the segment embedding indicate whether a token belongs to the first or second instruction in the pair. We use two unsupervised learning tasks for pre-training.

\begin{figure}[htb]
  \centering
  \includegraphics[width=0.95\linewidth]{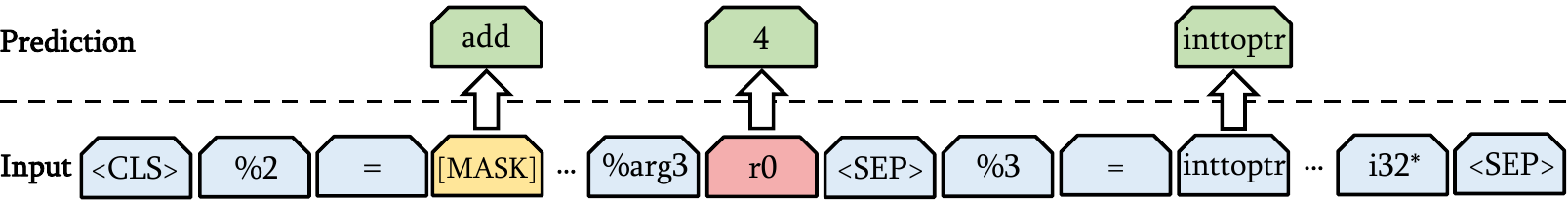}
  \caption{Masked Language Model (MLM).} 
  \vspace{-0.6ex}
  \label{fig:mlm}
\end{figure}

\emph{Task1: Masked Language Model.} We introduce an MLM task to help the model understand the internal structure details of LLVM-IR instructions. Figure \ref{fig:mlm} shows an example, we first randomly select 15\% of the tokens to replace: 70\% of them are replaced with [MASK], 15\% are replaced with random token, and 15\% remain unchanged. Then the model is required to predict the original token corresponding to the replaced position. The loss function uses Cross-Entropy loss as follows:
\begin{equation}
\footnotesize
\mathcal{L}_\text{MLM} = -\sum_{i \in M} \log p(y_i = \hat{y}_i), \hat{y}_i\in\{1, 2, \dots, |V|\}
\end{equation}
where $y$ represents the actual token, $\hat{y}$ represents the predicted token, $M$ is the masked token set and $|V|$ is the size of vocabulary.

\begin{figure}[htb]
  \centering
  \includegraphics[width=0.95\linewidth]{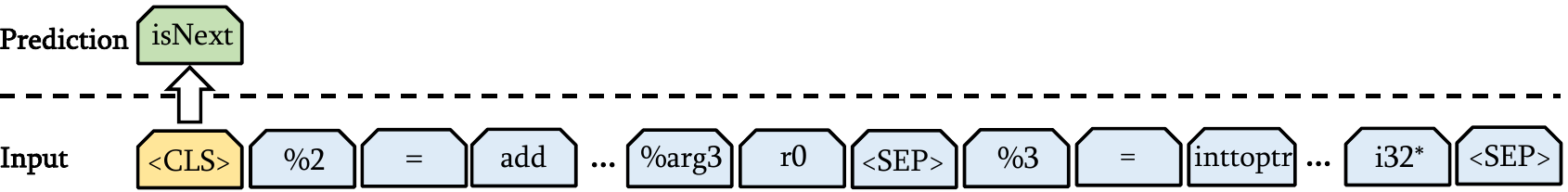}
  \caption{Next Sentence Prediction (NSP).} 
  \vspace{-0.6ex}
  \label{fig:nsp}
\end{figure}

\emph{Task2: Next Sentence Prediction.} We set up an NSP task to enable the model to capture control flow relationships between adjacent instructions. This task involves determining whether two instructions are control flow neighbors. As shown in Figure \ref{fig:nsp}, for an input instruction pair A+B, if B is the next instruction following A, it is regarded as a positive example (labeled as \texttt{isNext}), otherwise it is regarded as a negative example (labeled as \texttt{NotNext}). Positive and negative examples each constitute 50\% of the training data. We feed the final hidden state of the [CLS] token into the NSP head for binary classification. The loss function is computed using Cross-Entropy loss:
\begin{equation}
\footnotesize
\mathcal{L}_\text{NSP} = -\sum_{i \in D} \log p(y_i = \hat{y}_i)
\end{equation}
where $y$ represents the actual label, $\hat{y}$ represents the predicted result, and $D$ represents the training set.

The total loss function of the LLVM-IR language model is the combination of two loss functions:
\begin{equation}
\footnotesize
\mathcal{L}_\text{Total} = \mathcal{L}_\text{MLM} + \mathcal{L}_\text{NSP}
\end{equation}

After completing the pre-training of the LLVM-IR language model, we can input the instruction sequence into the model in units of basic blocks, to obtain the embedding at the basic block level.

\subsection{Graph Momentum Contrastive Learning} \label{sec:graph}

\noindent\textbf{Gated Graph Neural Networks.} 
After obtaining the embedding of the basic blocks, we use GGNN \cite{li2015gated} to transfer and aggregate the information of adjacent basic blocks on the CFG to learn the function embedding features. Specifically, the input CFG is represented as $G_i=\{V, E\}$, where the initial state of node $v \in V$ is a D-dimensional embedding, i.e., $\mathbf{h}_v^1 \in \mathbb{R}^D$. 
At each time step, each basic block node and its adjacent nodes transfer information, as follows:
\begin{equation}
\footnotesize
\mathbf{a}_v^t = A_{v_:}^T \left[ \mathbf{h}_1^{(t-1)T} \dots \mathbf{h}_{|v|}^{(t-1)T} \right] + \mathbf{b}
\end{equation}
where matrix $A_{v_:}^T$ is the transformation matrix of the adjacent node $v$, and ${a}_v^t$ represents the aggregate representation of node $v$ at step $v$.

When each node is fully updated, the information of all basic block nodes is used to calculate the output representation of the entire CFG, as follows:
\begin{equation}
\footnotesize
h_G = \tanh \left( \sum_{v \in V} \sigma \left( i(h_v^T) \right) \odot \tanh \left( j(h_v^T) \right) \right)
\end{equation}
where $i$ and $j$ represent two MLP, and $\sigma$ is the sigmoid function.

\begin{figure}[t]
  \centering
  \includegraphics[width=0.92\linewidth]{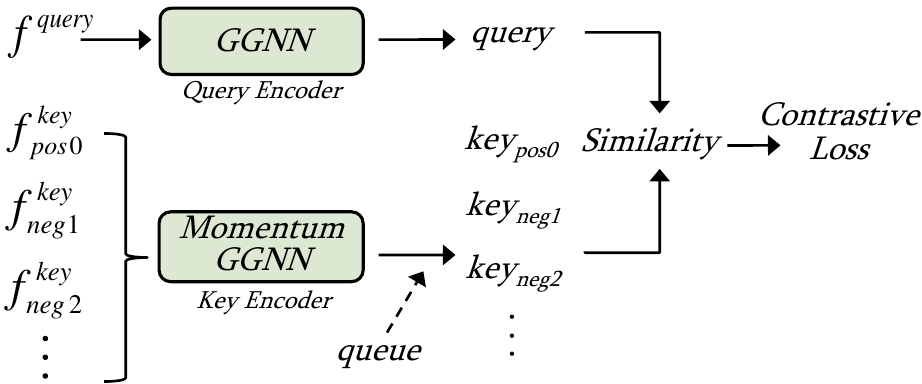}
  \vspace{-0.5ex}
  \caption{Framework of Momentum Contrast (MoCo).} 
  \vspace{-1.8ex}
  \label{fig:contrastive}
\end{figure}

\noindent\textbf{Momentum Contrastive Learning.} To bring similar functions (i.e., positive samples) closer to each other in the feature space, while pushing dissimilar functions (negative samples) further apart, we use contrastive learning for training. Specifically, as shown in Figure \ref{fig:contrastive}, given a function $f_{query}$, we sample one positive sample $f_{pos0}^{key}$ and $n$ negative samples $f_{neg1}^{key}$, $f_{neg2}^{key}$, ..., $f_{negn}^{key}$, which are encoded using two GGNN with shared weights. Existing research \cite{gao2021simcse} has shown that increasing the coverage of negative samples can improve contrastive learning performance. Therefore, we adopt the solution proposed in MoCo \cite{he2020momentum}, which maintains an embedding queue to facilitate training with larger batch sizes while minimizing memory consumption. However, due to the size of the queue, directly back-propagating to update the key encoder is challenging, so we use momentum updates instead:
\begin{equation}
\footnotesize
\theta_k \leftarrow m\theta_k + (1 - m)\theta_q
\end{equation}
where $\theta_q$ represents the parameters of the query encoder, $\theta_k$ represents the parameters of the key encoder, and $m$ is the momentum coefficient. We use InfoNCE as the loss function:
\begin{equation}
\footnotesize
\mathcal{L}_q = -\log \frac{\exp(q \cdot k_{pos0} / \tau)}{\sum_{i=1}^{n} \exp(q \cdot k_{negi} / \tau)}
\end{equation}
where $\tau$ is the temperature hyperparameter, $q$, $k_{pos0}$, $k_{negi}$ are the embeddings of $f_{query}$, $f_{pos0}^{key}$, and $f_{negi}^{key}$ respectively.

\subsection{Similarity Detection}

For a function pair <${f_1}$, ${f_2}$>, we determine their similarity based on the source file name and function name (i.e., whether they are compiled from the same source code), assigning a label of 1 for similar and 0 for dissimilar. Cosine distance is used to calculate the similarity between their embedding vectors <$\vec{f_1}$, $\vec{f_2}$>, with the formula shown below:
\begin{equation}
\footnotesize
\text{Similarity}(\vec{f_1}, \vec{f_2}) = \frac{\sum_{i=1}^{n} (\vec{f_1}[i] \cdot \vec{f_2}[i])}{\sqrt{\sum_{i=1}^{n} \vec{f_1}[i]^2} \cdot \sqrt{\sum_{i=1}^{n} \vec{f_2}[i]^2}}
\end{equation}
where $\vec{f}[i]$ represents \emph{i}-th component of vector $\vec{f}$.

\section{Experimental Setups}
\vspace{-0.6ex}
\subsection{Datasets} \label{sec:datasets}
\vspace{-0.3ex}
Referring to previous works \cite{marcelli2022machine, ding2019asm2vec}, we use the following 16 projects that are widely used in practice and related works, i.e. Binutils-2.34, Coreutils-8.32, ClamAV-0.102.0, Curl-7.67.0, Diffutils-3.7, Findutils-4.7.0, GMP-6.2.0, ImageMagick-7.0.10, Libmicrohttpd-0.9.71, LibTomCrypt-1.18.2, Nmap-7.80, OpenSSL-3.0, PuTTy-0.74, SQLite-3.34, Unrar-5.5.3, and Zlib-1.2.11.

We cross-compile these projects to obtain binary files in different compilation environments. Specifically, we use two compiler families (\texttt{GCC} and \texttt{Clang}), with four versions each, five optimization levels (\texttt{O0-O3}, \texttt{Os}), and five target architectures (\texttt{x86\_32}, \texttt{x86\_64}, \texttt{arm\_32}, \texttt{arm\_64}, \texttt{mips\_32}). We disable function inlining during compilation to avoid introducing noise. 
In total, we obtain 7,513 binaries. After completing the Data Pre-processing in Section \ref{sec:datapre}, they are devided into training and test sets by project for subsequent steps. The datasets details are provided in Appendix \ref{sec:dataset}.

\begin{figure}[t]
  \centering
  \includegraphics[width=0.92\linewidth]{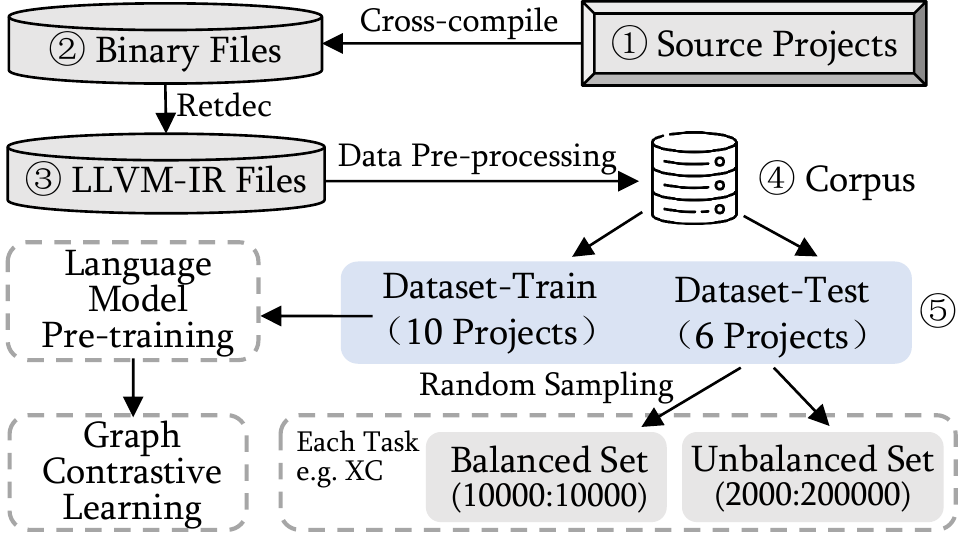}
  \vspace{-0.6ex}
  \caption{Dataset construction process.}
  \vspace{-2.3ex}
  \label{fig:dataset}
\end{figure}

To validate the capability of \ModName~in complex compilation environments, we identify seven different tasks, i.e. \texttt{XC} (cross-compiler), \texttt{XO} (cross-optimization), \texttt{XA} (cross-architecture), and their combinations. For example, the task \texttt{XO+XA} represents that the function pairs have different optimization levels and architectures but use the same compiler. 
As in experimental setup of previous works \cite{pei2020trex,li2021palmtree}, we construct a balanced testset for each task, consisting of 10K positive function pairs and 10K negative function pairs, as well as an unbalanced testset consisting of 2K positive function pairs and 200K negative function pairs.
The complete dataset construction process is shown in Figure \ref{fig:dataset}.

\vspace{-0.6ex}
\subsection{Baselines}
\vspace{-0.8ex}

We compare \ModName~to seven baselines:

\textbf{Zeek} \cite{shalev2018binary}, performs dataflow analysis on the at the basic-block level and computes stands, and train a two-layer fully CNN to learn the similarity task.

\textbf{Gemini} \cite{xu2017neural}, extracts manually crafted features for each basic block to construct ACFG, and uses GNN for representation.

\textbf{Asm2Vec} \cite{ding2019asm2vec}, utilizes random walks on the CFG to sample instruction sequences, and then uses the unsupervised learning model PV-DM to learn assembly functions representation.

\textbf{SAFE} \cite{massarelli2019safe}, uses the word2vec model to generate assembly instruction embeddings, and adopts RNN with attention mechanism to generate function embeddings.

\textbf{GMN} \cite{li2019graph}, directly uses a graph matching network based on a control-flow graph to calculate the similarity of two functions.

\textbf{Trex} \cite{pei2020trex}, uses a transfer learning-based framework to generate function embeddings based on micro-traces of function.

\textbf{PalmTree} \cite{li2021palmtree}, adds pre-training tasks related to code structure to improve the BERT model, and performs self-supervised pre-training on a large-scale unlabeled binary corpus to generate instruction embeddings.

\subsection{Metrics}
\vspace{-0.8ex}
Following previous researches \cite{marcelli2022machine, pei2020trex}, we conduct experimental evaluation in two scenarios:

\noindent\textbf{One-to-one Comparison}. Given a pair of binary functions, return their similarity score to determine whether they are similar. It can be regarded as a binary classification task, and we use the Area Under the Curve (AUC) of the Receiver Operating Characteristic (ROC) curve as the evaluation metric, which is a comprehensive measure that incorporates all possible classification thresholds

\noindent\textbf{One-to-many Search}. Given a query binary function \emph{f} $\in$ \emph{F}, and a binay function pool \emph{P}, where \emph{P} contains a function $f^{gt}$ that is similar to \emph{f}, and |\emph{P}|-1 dissimilar functions, the objective is to retrieve the Top-k functions from the pool \emph{P} ranked by similarity to \emph{f}. The rank of each retrieved function $f_i$ is denoted by Rank$_{f_i}$.
The indicator function $g$ and Recall@\emph{k} are defined as follows:
\begin{equation}
\footnotesize
g(x) = \begin{cases}
1 \quad \text{if } x = True \\
0 \quad \text{if } x = False 
\end{cases} 
\end{equation}
\begin{equation}
\footnotesize
\text{Recall}@k = \frac{1}{|F|} \sum_{i=1}^{|F|} g(\text{Rank}_{f_i^{gt}} \leq k)
\end{equation}
where \emph{F} represents the total number of queries. While Recall@\emph{k} emphasizes the retrieval coverage rate, MRR places greater emphasis on the order and positional relationship within the rank list. It is calculated as:
\vspace{-0.6ex}
\begin{equation}
\footnotesize
\text{MRR} = \frac{1}{|F|} \sum_{i=1}^{|F|} \frac{1}{\text{Rank}_{f_i^{gt}}}
\end{equation}

\begin{table*}[htbp]
\centering
\caption{AUC score of One-to-one comparison on balanced and unbalanced set.}
\setlength{\tabcolsep}{0.9mm}
\scalebox{0.68}{
\begin{threeparttable}
    \begin{tabular}{lcccccccc|cccccccc}
    \toprule \hline
    \multirow{2}{*}{} & \multicolumn{7}{c}{Balanced Set} & \multirow{2}{*}{Avg.} & \multicolumn{7}{c}{Unbalanced Set} & \multirow{2}{*}{Avg.} \\ 
    \cmidrule(r){2-8}  \cmidrule(r){10-16} 
    & XC & XO & XA & XC+XO & XO+XA & XC+XA & XC+XA+XO & & XC & XO & XA & XC+XO & XO+XA & XC+XA & XC+XA+XO & \\ 
    \midrule
    Zeek & 0.955 & 0.924 & 0.949 & 0.920 & 0.917 & 0.928 & 0.912 & 0.929 
    & 0.956 & 0.924 & 0.947 & 0.914 & 0.913 & 0.922  & 0.912 & 0.927 \\
    Gemini & 0.954 & 0.923 & 0.947 & 0.929 & 0.923 & 0.935 & 0.924 & 0.934 
    & 0.955 & 0.930 & 0.933 & 0.929 & 0.928 & 0.920  & 0.921 & 0.931  \\
    Asm2Vec\tnote{1} & 0.721 & 0.684 & - & 0.632 & - & - & - & - 
    & 0.733 & 0.680 & - & 0.642 & - & - & - & -  \\
    SAFE   & 0.918 & 0.893 & 0.909 & 0.895 & 0.888 & 0.893 & 0.881 & 0.897 
    & 0.918 & 0.891 & 0.902 & 0.894 & 0.889 & 0.893 & 0.882 & 0.896 \\
    GMN   & 0.891 & 0.816 & 0.669 & 0.787 & 0.648 & 0.633 & 0.605 & 0.721
    & 0.889 & 0.817 & 0.661 & 0.786 & 0.652 & 0.643  & 0.614 & 0.723 \\
    Trex  & 0.970 & 0.948 & 0.939 & 0.941 & 0.931 & 0.926 & 0.914 & 0.938
    & 0.970 & 0.946 & 0.943 & 0.942 & \textbf{0.932} & 0.923  & 0.914 & 0.939 \\
    PalmTree\tnote{1} & 0.959 & 0.954 & - & 0.938 & - & - & - & - 
    & 0.957 & 0.950 & - & 0.935 & - & - & - & -  \\
    \midrule
    \rowcolor{gray!30}
    \textbf{IRBinDiff}  & \textbf{0.986} & \textbf{0.968} & \textbf{0.959} & \textbf{0.970} & \textbf{0.931} & \textbf{0.943} & \textbf{0.935} & \textbf{0.956} & 
    \textbf{0.981} & \textbf{0.961} & \textbf{0.946} & \textbf{0.957} & 0.922 & \textbf{0.929} & \textbf{0.927} & \textbf{0.946} \\ 
    w/o Norm & 0.981 & 0.960 & 0.949 & 0.964 & 0.917 & 0.935 & 0.925 & 0.947 
    & 0.977 & 0.951 & 0.940 & 0.951 & 0.906 & 0.920 & 0.909 & 0.936 \\
    w/o PLM & 0.975 & 0.950 & 0.939 & 0.951 & 0.906 & 0.915 & 0.896 & 0.933 
    & 0.968 & 0.945 & 0.929 & 0.934 & 0.891 & 0.901 & 0.882 & 0.921 \\
    w/o Graph & 0.964 & 0.920 & 0.790 & 0.925 & 0.753 & 0.782 & 0.784 & 0.845 
    & 0.967 & 0.923 & 0.790 & 0.921 & 0.764 & 0.785 & 0.785 & 0.848 \\
    \hline \bottomrule 
    \end{tabular}
        \begin{tablenotes}
            \item[1] Asm2Vec and PalmTree can not support cross-architecture (XA) tasks.
        \end{tablenotes}
\end{threeparttable}
}
\label{tab:one2one_results}
\end{table*}

\vspace{-1.8ex}
\subsection{Implementation}
\vspace{-0.3ex}
Our experimental environment is a machine running on Ubuntu 20.04 OS, equipped with 10 * NVIDIA GeForce RTX 3090 GPU and a 48-core Intel Xeon Gold 5220R CPU. We use Python 3.8 with PyTorch 1.11.0 to implement \ModName~ and all experiments. 

In the Data Pre-processing stage, we filter out all functions with less than 5 basic blocks, tokenize the simplified and normalized instructions, and generate a vocabulary size of 26,608.

In the Language Model Pre-training stage, we use the Adam optimizer with an initial learning rate of 3e-5, 4 hidden layers, 128 hidden dimension, 8 multi-head attention, 256 training batches, and 3 epochs. The basic block embedding generated is represented by a 128-dimensional vector.

In the Graph Contrastive Learning stage, we divide the training and validation set in a ratio of 8.5:1.5, use the Adam optimizer with an initial learning rate of 1e-4 and a weight decay value of 5e-4, and the loss function is InfoNCE. The model has 10 encoding layers with 256 units per layer, 256 training batches, and 10 epochs. The momentum coefficient is 0.999, and the embedding queue length is 8,192.
The function embedding generated is represented by a 256-dimensional vector.

\section{Experimental Results}

\subsection{One-to-one Comparison}
Function similarity detection in One-to-one scenario is the most basic and relatively simple task in BCSD. Referring to the settings of previous work \cite{li2021palmtree,pei2020trex,marcelli2022machine}, we use the balanced and unbalanced testset mentioned in Section \ref{sec:datasets} to evaluate. Table \ref{tab:one2one_results} shows the AUC score of \ModName~and other baselines.

Overall, \ModName~outperforms other methods in XC, XO, XA and their combined tasks on balanced testset. Specifically, \ModName's AUC scores in seven tasks are improved by 9.4\%, 11.7\%, 10.7\%, 14.5\%, 10.2\%, 11.8\% and 13.3\% on average compared to all baseline methods. For example, in the most difficult XC+XO+XA task, \ModName~gets 0.935, the second highest Gemini gets 0.924, while GMN and SAFE perform significantly worse, only 0.605 and 0.881. It is worth noting that the Asm2Vec and PalmTree only support the x86 instruction set architecture, so they cannot be compared on cross-architecture (XA) tasks.

For the unbalanced testset, \ModName~also performs well despite the more challenging data distribution. On the XO+XA task, although \ModName~fell slightly behind Gemini and Trex by a margin of 0.006 and 0.010, it still exhibits robust overall performance. Compared to Zeek, Gemini, SAFE, GMN, and Trex, the average AUC scores of \ModName~are improved by 2.1\%, 1.6\%, 5.6\%, 30.8\%, and 0.8\%. In conclusion, the superior and robust performance of \ModName~can be attributed to its ability to lift binary code to a unified LLVM-IR representation, effectively mitigating the discrepancies caused by different compilation options.

\subsection{One-to-many Search}

\begin{figure}[t]
    \centering
    \begin{subfigure}[]{0.494\linewidth}
        \centering
        \includegraphics[width=\linewidth]{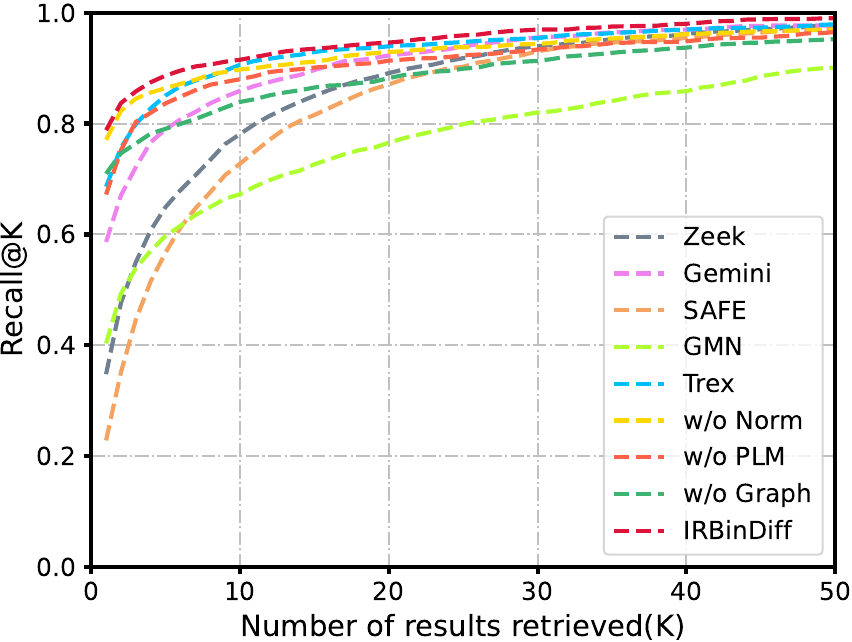}
        \caption{XO}
        \label{fig:XO}
    \end{subfigure}
    \begin{subfigure}[]{0.494\linewidth}
        \centering
        \includegraphics[width=\linewidth]{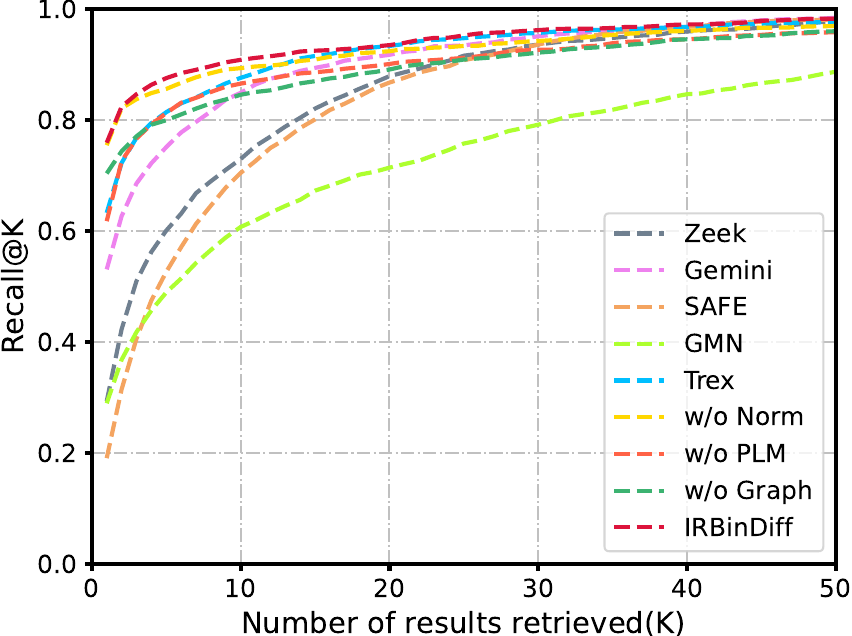}
        \caption{XO+XC}
        \label{fig:XO_XC}
    \end{subfigure}
    \begin{subfigure}[b]{0.494\linewidth}
        \centering
        \includegraphics[width=\linewidth]{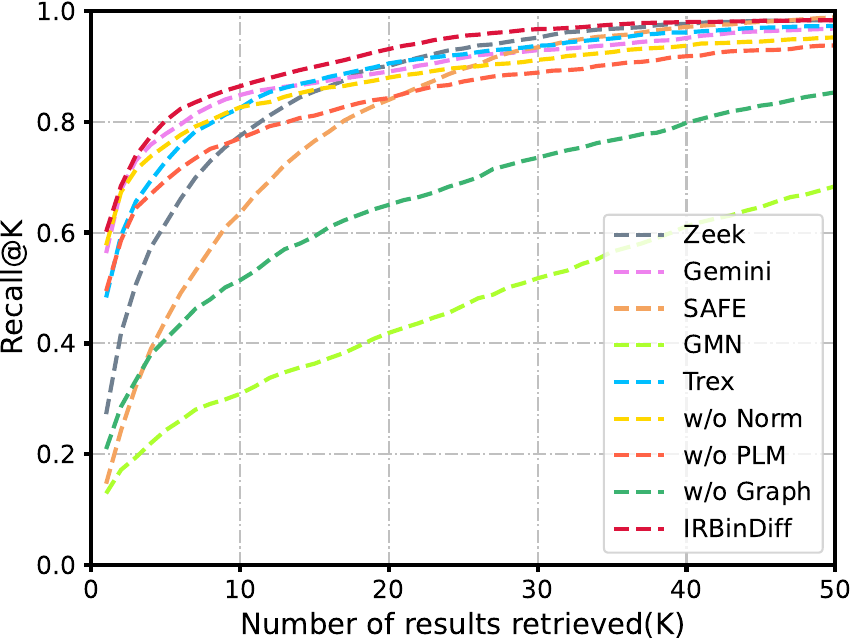}
        \caption{XC+XA}
        \label{fig:XC_XA}
    \end{subfigure}
    \begin{subfigure}[b]{0.494\linewidth}
        \centering
        \includegraphics[width=\linewidth]{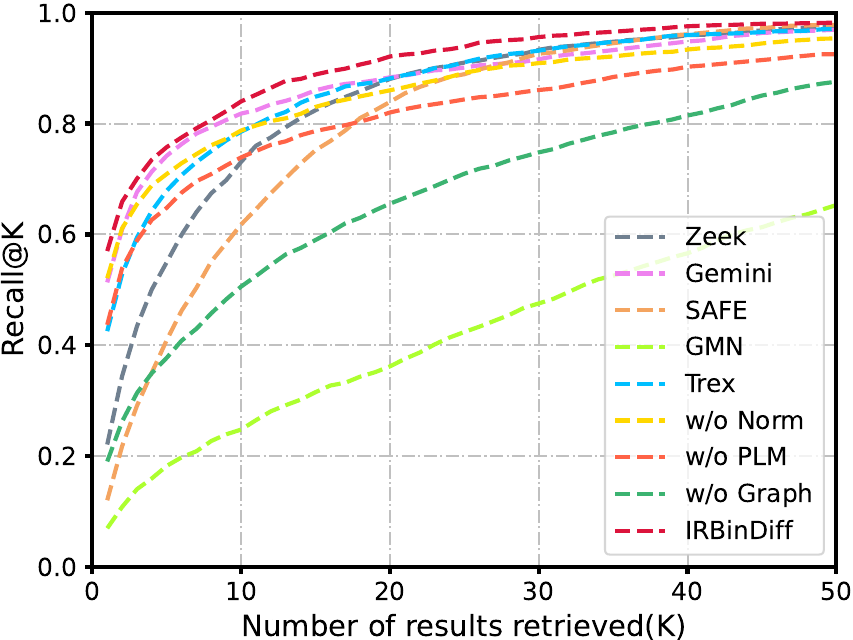}
        \caption{XC+XO+XA}
        \label{fig:XC_XO_XA}
    \end{subfigure}
    \vspace{-3.0ex}
    \caption{Recall@K score at different K values of One-to-many search on unbalanced set.}
    \vspace{-2.0ex}
    \label{fig:recallK}
\end{figure}

One-to-many search presents greater challenges but is a highly practical scenario, for example, in large-scale firmware analysis and vulnerability detection, it is usually necessary to retrieve similar functions from large-scale candidate functions \cite{zhao2022large}. To evaluate this, we use the unbalanced testset described in Section \ref{sec:datasets}. Specifically, for each query function, we construct a candidate function pool of size 101, consisting of 1 similar and 100 dissimilar functions.

Table \ref{tab:one2many_results} presents Recall@1 and MRR scores of \ModName~and baselines. Notably, \ModName~achieves an average relative improvement in Recall@1 of between 75.7\% and 185\% across seven tasks compared to seven baseline methods.
Especially in the most challenging XC+XA+XO tasks, \ModName relatively improves Recall@1 score by up to 185\%. For MRR scores, \ModName~only trails Gemini in XA by a mere 0.004 points.
Overall, compared to One-on-one comparison, \ModName~demonstrates more significant performance advantages in the One-to-many search scenario. This performance boost is primarily attributed to the introduction of momentum contrastive learning, which effectively enhances the model’s ability to distinguish subtle functional differences in large-scale candidate pools, enabling it to accurately identify similar functions.

Figure \ref{fig:recallK} reports the Recall@K scores of each method under different K values. As shown, from task XO to XC+XO+XA, as the difference in compilation options increases, the search performance of each BCSD method shows a downward trend. However, in each task, as the K value increases, \ModName~consistently surpasses the baseline methods, with its Recall@K curve remaining above the others, demonstrating higher accuracy at every recall level.

\begin{table*}[htbp]
\centering
\caption{Recall@1 and MRR score of One-to-many search on unbalanced set (Poolsize = 101).}
\setlength{\tabcolsep}{0.9mm}
\scalebox{0.68}{
\begin{threeparttable}
    \begin{tabular}{lcccccccc|cccccccc}
    \toprule \hline
    \multirow{2}{*}{} & \multicolumn{7}{c}{Recall@1} & \multirow{2}{*}{Avg.} & \multicolumn{7}{c}{MRR} & \multirow{2}{*}{Avg.} \\ 
    \cmidrule(r){2-8}  \cmidrule(r){10-16} 
    & XC & XO & XA & XC+XO & XO+XA & XC+XA & XC+XA+XO &  & XC & XO & XA & XC+XO & XO+XA & XC+XA & XC+XA+XO &  \\ 
    \midrule
    Zeek & 0.447 & 0.348 & 0.293 & 0.293 & 0.274 & 0.272 & 0.221 & 0.307
    & 0.595 & 0.487 & 0.465 & 0.439 & 0.427 & 0.433 & 0.377 & 0.460 \\
    Gemini & 0.775 & 0.586 & 0.644 & 0.531 & 0.527 & 0.563 & 0.513 & 0.591
    & 0.829 & 0.677 & \textbf{0.742} & 0.634 & 0.634 & 0.672 & 0.620 & 0.687  \\
    Asm2Vec\tnote{1} & 0.321 & 0.254 & - & 0.174 & - & - & - & -
    & 0.405 & 0.324 & - & 0.248 & - & - & - & -  \\
    SAFE   & 0.306 & 0.228 & 0.149 & 0.190 & 0.138 & 0.147 & 0.120 & 0.183
    & 0.461 & 0.384 & 0.308 & 0.349 & 0.292 & 0.294 & 0.268 & 0.337 \\
    GMN   & 0.573 & 0.403 & 0.179 & 0.289 & 0.101 & 0.129 & 0.070 & 0.249 
    & 0.659 & 0.499 & 0.244 & 0.391 & 0.173 & 0.198 & 0.139 & 0.329 \\
    Trex  & 0.833 & 0.686 & 0.579 & 0.633 & 0.525 & 0.483 & 0.425 & 0.594 
    & 0.878 & 0.759 & 0.682 & 0.717 & 0.631 & 0.597 & 0.542 & 0.686 \\
    PalmTree\tnote{1} & 0.779 & 0.645 & - & 0.597 & - & - & - & -
    & 0.835 & 0.713 & - & 0.681 & - & - & - & -  \\
    \midrule
    \rowcolor{gray!30}
    \textbf{IRBinDiff}  & \textbf{0.870} & \textbf{0.788} & \textbf{0.661} & \textbf{0.760} & \textbf{0.550} & \textbf{0.602} & \textbf{0.570} & \textbf{0.686}
    & \textbf{0.906} & \textbf{0.833} & 0.738 & \textbf{0.814} & \textbf{0.640} & \textbf{0.684} & \textbf{0.656} & \textbf{0.753} \\ 
    w/o Norm & 0.863 & 0.771 & 0.643 & 0.755 & 0.523 & 0.577 & 0.521 & 0.665 
    & 0.899 & 0.817 & 0.723 & 0.807 & 0.610 & 0.664 & 0.612 & 0.733 \\
    w/o PLM & 0.717 & 0.672 & 0.564 & 0.618 & 0.470 & 0.495 & 0.437 & 0.568
    & 0.803 & 0.747 & 0.658 & 0.708 & 0.566 & 0.590 & 0.541 & 0.659 \\
    w/o Graph & 0.846 & 0.710 & 0.233 & 0.704 & 0.188 & 0.210 & 0.191 & 0.440
    & 0.875 & 0.752 & 0.331 & 0.751 & 0.287 & 0.313 & 0.294 & 0.515  \\
    \hline \bottomrule
    \end{tabular}
        \begin{tablenotes}
            \item[1] Asm2Vec and PalmTree can not support cross-architecture (XA) tasks.
        \end{tablenotes}
\end{threeparttable}
}
\label{tab:one2many_results}
\end{table*}

\subsection{Ablation Study}
We conduct an ablation study by comparing \ModName~with the following variants to better understand the contribution of each component:
\begin{itemize}[left=0.2cm] 
    \setlength{\itemsep}{0pt}  
    \vspace{-0.5ex}
    \item \textbf{w/o Norm}: Removing instruction normalization process in the first part of Section \ref{sec:languagemodel}.
    \item \textbf{w/o PLM}: Removing the pre-trained language model described in Section \ref{sec:languagemodel}.
    \item \textbf{w/o Graph}: Removing the graph momentum contrastive learning described in Section \ref{sec:graph}.
\end{itemize}
\vspace{-0.5ex}
As shown in Table \ref{tab:one2one_results} and Table \ref{tab:one2many_results}, removing any component negatively impacts the overall performance of \ModName.
Among them, instruction normalization component has the least impact on performance, with the w/o Norm variant causing an average performance drop of only 0.010 and 0.011 points in the One-to-one and One-to-many scenarios, respectively, but its contribution to reducing training resources and time overhead is extremely substantial.

In contrast, the w/o PLM variant, which removes the pre-trained language model, results in an average drop of 0.025 and 0.106 points in the two scenarios. This highlights the significance of LLVM-IR’s higher-level semantic abstraction and the fine-grained semantic understanding provided by the pre-trained language model.
Finally, the graph momentum contrastive learning component proves to be the most critical in \ModName. The w/o Graph variant, which excludes this component, leads to a significant performance drop of 0.101 and 0.242 points, particularly in tasks involving XA.

These conclusions are further supported by the Recall@K curves for different K values, as shown in Figure \ref{fig:recallK}, which underscore the essential role of all three components in the overall performance.

\section{Conclusion}
In this paper, we present \ModName, a novel graph momentum contrastive learning approach based on intermediate representations (LLVM-IR) for binary code similarity detection. Our approach successfully tackles the challenges of complex compilation options and large-scale candidate function retrieval in real-world applications, consistently outperforming existing leading solutions across various experimental tasks and scenarios. In conclusion, \ModName~offers a robust and practical research solution for binary code similarity detection.

\section*{Limitations}
Although our proposed \ModName~performs well in experiments, it still has certain limitations that need to be addressed in subsequent studies:

First, \ModName~currently does not take into account code obfuscation. Code obfuscation is commonly used to evade static analysis and can significantly alter the logic and structure of a program, making it more challenging to analyze. We consider code obfuscation an orthogonal issue, and future advancements in handling obfuscation techniques could complement the existing capabilities of \ModName.

Secondly, our work uses sixteen engineering projects to construct binary code datasets, which may not encompass all possible application scenarios. Nevertheless, these projects are widely utilized in both related research and practical applications and can fully reflect the diversity of software in the real world, thus alleviating potential concerns about the generalizability of our method.

Finally, \ModName's overall performance is partly dependent on the generation and preprocessing of LLVM-IR, which remains a computationally intensive task. This is particularly evident when processing large-scale binary files, where the efficiency of the Retdec decompilation tool may present a bottleneck.

\section*{Ethical Considerations}
We state that this research strictly adheres to the ethical and scientific research principles in the field of computer science.

\bibliography{main}

\begin{figure*}[h]
  \centering
  \includegraphics[width=0.92\linewidth]{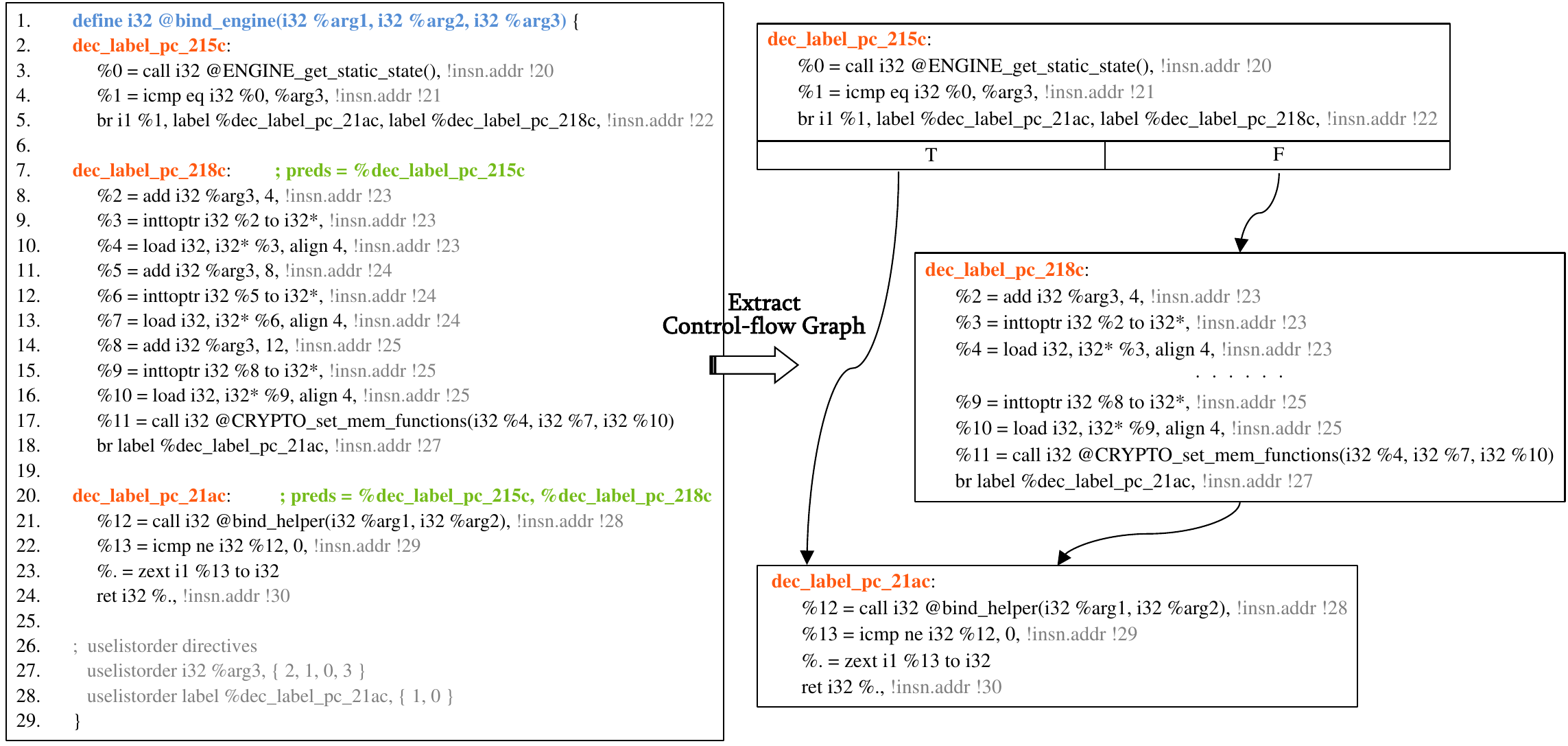}
  \caption{The LLVM-IR readable form (Left) and the corresponding control-flow graph form (Right) of the \texttt{bind\_engine} function in the \texttt{OpenSSL} project.} 
  \label{fig:cfg}
\end{figure*}

\appendix

\begin{table*}[h]
    \centering
    \caption{Examples of LLVM-IR instruction tokenization and normalization.}
    \label{tab:rule}
    \setlength{\tabcolsep}{1.02mm}
    \scalebox{0.78}{
    \begin{threeparttable}
    \begin{tabular}{ccc}
        \toprule \hline
        \textbf{Rule ID} & \textbf{Original Instruction} & \textbf{Tokenized and Normalized Instruction}  \\
        \midrule
        (1) & br i1 \%22, label \textcolor{customred}{\%dec\_label\_pc\_41d34}, label \textcolor{customred}{\%dec\_label\_pc\_41d28} & br i1 \%22 , label \textcolor{customgreen}{<label>} , label \textcolor{customgreen}{<label>} \\
        (2) & \%54 = load i32, i32* \textcolor{customred}{@global\_var\_136b14}, align 4 & \%54 = load i32 , i32* \textcolor{customgreen}{<global>} , align 4 \\
        (3) & call void @\_\_asm\_fcmpe(float \%6, float \textcolor{customred}{0x43F0000000000000}) & call void @ asm fcmpe ( float \%6 , float \textcolor{customgreen}{<Positive>} )  \\
        (3) & \%278 = add nsw i32 \%277, \textcolor{customred}{-630} & \%278 = add nsw i32 \%277 , \textcolor{customgreen}{<Negative>}  \\
        (4) & store i32 \textcolor{customred}{4325376}, i32* \%s1.0.reg2mem & store i32 \textcolor{customgreen}{<Address>} , i32* \%s1 0 reg2mem \\
        (5) & store i8* \%65, i8** \textcolor{customred}{\%storemerge518}.reg2mem & store i8* \%65 , i8** \textcolor{customgreen}{\%storemerge} reg2mem \\
        (5) & \textcolor{customred}{\%or.cond10} = or i1 \%brmerge, \textcolor{customred}{\%or.cond4} & \textcolor{customgreen}{\%or cond} = or i1 \%brmerge , \textcolor{customgreen}{\%or cond} \\
        \hline \bottomrule
    \end{tabular}
    \end{threeparttable}
    }
\end{table*}

\section{Data Pre-processing Sumpplements} \label{sec:datapreappendix}

\subsection{Extract Control-flow Graph Sumpplements} \label{sec:cfgappendix}
LLVM-IR is a language with clear semantics, and the \texttt{.ll} file generated by decompiling binary files is its readable form. The example function shown in Figure \ref{fig:cfg} is obtained from the \texttt{bind\_engine} function in the \texttt{afalg.so} file of the \texttt{OpenSSL} project. From the example, it can be observed that LLVM-IR resembles the RISC instruction set, supporting simple operations such as addition, subtraction, branching, and comparison, with instructions presented in three-address format.

\noindent\textbf{Function Boundary Division.} As shown in Figure \ref{fig:cfg}, each function in LLVM-IR begins with the keyword '\texttt{define}'. We sequentially traverse the LLVM-IR statements, using regular expressions to identify '\texttt{define}', and add each subsequent statement to the current function block until we encounter the next '\texttt{define}' or reach the end of the file, thereby marking the end of the function block.

\noindent\textbf{Basic Block Division.} LLVM provides the \texttt{opt} command to obtain a graphical CFG, but we need to extract it in text form. Therefore, we employ regular expression matching to identify the identifier indicating basic block call relationships from the function blocks to construct the CFG.

As observed in Figure \ref{fig:cfg}, the identifier "\texttt{dec\_label\_pc\_xxxx:}" indicates the start of a basic block, while "\texttt{;preds= \%dec\_label\_pc\_xxxx}" indicates the predecessors of the current basic block. 
For instance, in line 20, "\texttt{dec\_label\_pc\_ 21ac: ;preds = \%dec\_label\_pc\_215c, \%dec\_ label\_pc\_218c}" indicates that the basic blocks "\texttt{dec\_label\_pc\_215c}" and "\texttt{dec\_label\_pc\_218 c}" are predecessors of "\texttt{dec\_label\_pc\_21ac}".

Therefore, within each function block, we scan each statement, using regular expressions to identify "\texttt{dec\_label\_pc\_xxxx:}". The numbered identifiers are stored in a list as nodes of the CFG. Subsequently, each statement following the identifier is added to the current basic block until we encounter the next identifier or reach the end of the function block, thus concluding the current basic block. 

\noindent\textbf{Control-flow Graph Extraction.}
After completing the division of all basic blocks, we check whether the basic block contains the "\texttt{;preds =}" identifier. If present, we locate the "\texttt{dec\_label\_pc\_xxxx}" that follows this identifier in the list and form a pair with the current basic block identifier, representing an edge in the CFG, which is stored in another list. Once the extraction of nodes and edges is complete, the extraction of the CFG is also finalized.

\subsection{Instruction Simplification Sumpplements} \label{sec:simplifyappendix}

In order to prune redundant instructions while preserving essential ones, we empirically design the following instruction simplification strategies:

\textbf{(1) Basic Block Identifiers}: In LLVM-IR, the identifier "\texttt{dec\_label\_pc\_xxxx:}" and "\texttt{;preds = \%dec\_label\_pc\_xxxx}" indicate the start of a basic block and its predecessors, respectively, without any actual semantics. Since we have already divided basic block boundaries and extracted the CFG in Appendix \ref{sec:cfgappendix}, retaining these identifiers is unnecessary, so we remove them.

\textbf{(2) Address Information}: The statement "\texttt{, !insn.addr !xx}" following each instruction indicates its address. As the same instruction can have different addresses in different files, which may disrupt the model's semantic learning and lack inherent meaning, we delete it.

\textbf{(3) Memory Order Directives}: As shown in Figure \ref{fig:cfg}, the function body concludes with the list order directive "\texttt{; uselistorder directives}", which appears after the terminator of the last basic block and encodes the memory order for each usage list. These are not actual instructions and do not affect the semantics of the IR, so we delete them.

\section{Language Model Pre-training Sumpplements} \label{sec:pretrainappendix}

To provide a more intuitive illustration of our process for instruction tokenization and normalization, we present examples in Table \ref{tab:rule}, where the "Rule ID" column corresponds to the normalization rule numbers discussed in Section \ref{sec:languagemodel}.

\begin{figure*}[h]
  \centering
  \includegraphics[width=0.92\linewidth]{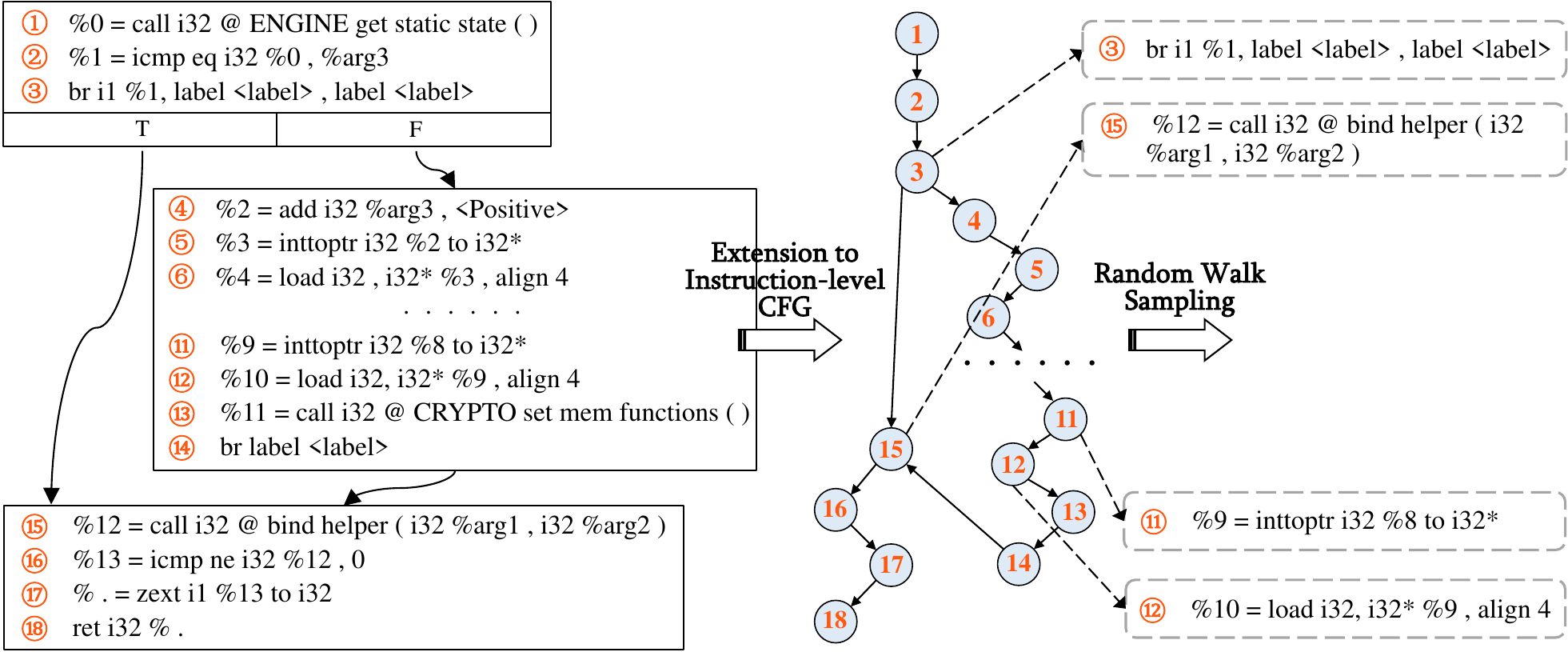}
  \caption{Example of random walk sampling instruction pairs.} 
  \label{fig:randomwalk}
\end{figure*}

Figure \ref{fig:randomwalk} presents an example of random walk sampling instruction pairs. Specifically, the original CFG consists of basic blocks and jump control-flow, where the nodes represent basic blocks, and the edges represent jump control-flow. By splitting basic blocks into individual instructions and adding sequential control-flow between them, the CFG can be extended to the instruction-level CFG, where the nodes represent single instructions and the edges represent either jump or sequential control-flow. Subsequently, we can perform random walk sampling on the instruction-level CFG, as described in Section \ref{sec:languagemodel}.

\section{Dataset Statistics} \label{sec:dataset}

\begin{table*}[htb]
    \centering
    \caption{Statistics of training and test datasets.}
    \label{tab:dataset}
    \setlength{\tabcolsep}{1.1mm}
    \scalebox{0.78}{
    \begin{threeparttable}
    \begin{tabular}{c|l|l|rrrrr}
        \toprule \hline
        \textbf{Dataset} & \textbf{Project} & \textbf{Version} & \textbf{\# Binaries} & \textbf{\# Functions} & \textbf{\# Filtered Functions} & \textbf{\# Basic Blocks} & \textbf{\# Instructions} \\
        \midrule
        \multirow{10}{*}{Train} & Binutils & 2.34 & 357 & 334,770 & 166,903 & 4,328,352 & 38,544,947 \\
        & Coreutils & 8.32 & 2,772 & 374,190 & 121,923 & 2,634,646 & 24,981,722 \\
        & Curl & 7.67.0 & 178 & 156,410 & 66,204 & 1,536,614 & 11,675,152 \\
        & Diffutils & 3.7 & 103 & 19,824 & 7,582 & 199,345 & 1,941,618 \\
        & Findutils & 4.7.0 & 103 & 27,575 & 10,908 & 241,597 & 2,162,108 \\
        & ImageMagick & 7.0.10 & 16 & 58,156 & 28,355 & 908,648 & 9,284,354 \\
        & Libmicrohttpd & 0.9.71 & 16 & 3,811 & 1,108 & 19,215 & 136,731 \\
        & LibTomCrypt & 1.18.2 & 16 & 17,968 & 4,201 & 74,677 & 808,328 \\
        & Unrar & 5.5.3 & 174 & 109,951 & 43,760 & 778,167 & 6,840,361 \\
        & Zlib & 1.2.11 & 735 & 71,677 & 40,951 & 804,488 & 8,329,255 \\
        \midrule
        \multirow{6}{*}{Test} & ClamAV & 0.102.0 & 812 & 2,226,318 & 621,533 & 10,125,334 & 86,078,874 \\
        & GMP & 6.2.0 & 16 & 16,246 & 7,111 & 168,892 & 1,596,924 \\
        & Nmap & 7.80 & 522 & 1,042,488 & 293,428 & 5,064,057 & 39,753,253 \\
        & OpenSSL & 3.0 & 1,602 & 1,901,218 & 478,963 & 6,693,744 & 56,906,444 \\
        & PuTTy & 0.74 & 84 & 65,826 & 21,625 & 379,777 & 4,329,182 \\
        & SQLite & 3.34.0 & 16 & 27,824 & 14,932 & 323,016 & 2,739,910 \\
        \midrule
        \rowcolor{gray!30}
        Total & \# 16 & - & 7,513 & 6,454,252 & 1,929,487 & 34,281,169 & 296,109,163 \\
        \hline \bottomrule
    \end{tabular}
    \end{threeparttable}
    }
\end{table*}

The statistics of the training and test datasets are shown in Table \ref{tab:dataset}.

\section{One-to-many Search Supplements} \label{sec:recall1050}

\begin{table*}[htbp]
\centering
\caption{Recall@10 and Recall@50 score of One-to-many search on unbalanced set (Poolsize = 101).}
\setlength{\tabcolsep}{0.9mm}
\scalebox{0.70}{
\begin{threeparttable}
    \begin{tabular}{lcccccccc|cccccccc}
    \toprule \hline
    \multirow{2}{*}{} & \multicolumn{7}{c}{Recall@10} & \multirow{2}{*}{Avg.} & \multicolumn{7}{c}{Recall@50} & \multirow{2}{*}{Avg.} \\ 
    \cmidrule(r){2-8}  \cmidrule(r){10-16} 
    & XC & XO & XA & XC+XO & XO+XA & XC+XA & XC+XA+XO &  & XC & XO & XA & XC+XO & XO+XA & XC+XA & XC+XA+XO &  \\ 
    \midrule
    Zeek & 0.875 & 0.781 & 0.835 & 0.730 & 0.757 & 0.774 & 0.731 & 0.783
    & 0.984 & 0.973 & 0.991 & 0.972 & 0.978 & 0.981  & 0.970 & 0.978 \\
    Gemini & 0.930 & 0.859 & \textbf{0.914} & 0.849 & 0.823 & 0.855 & 0.810 & 0.863 
    & 0.986 & 0.981 & 0.989 & 0.982 & 0.977 & 0.970 & 0.969 & 0.979  \\
    Asm2Vec\tnote{1} & 0.574 & 0.445 & - & 0.376 & - & - & - & - 
    & 0.841 & 0.773 & - & 0.704 & - & - & - & -  \\
    SAFE   & 0.809 & 0.727 & 0.683 & 0.705 & 0.653 & 0.635 & 0.618 & 0.690 
    & 0.985 & 0.972 & 0.991 & 0.982 & 0.980 & \textbf{0.988} & 0.978 & 0.982 \\
    GMN   & 0.825 & 0.672 & 0.348 & 0.607 & 0.299 & 0.309 & 0.248 & 0.473
    & 0.956 & 0.901 & 0.701 & 0.887 & 0.704 & 0.683 & 0.653 & 0.784 \\
    Trex  & 0.956 & 0.906 & 0.881 & 0.876 & 0.842 & 0.824 & 0.785 & 0.867 
    & 0.990 & 0.986 & 0.984 & 0.979 & 0.978 & 0.973 & 0.969 & 0.981 \\
    PalmTree\tnote{1} & 0.928 & 0.865 & - & 0.855 & - & - & - & -
    & 0.986 & 0.973 & - & 0.980 & - & - & - & -  \\
    \midrule
    \rowcolor{gray!30}
    \textbf{IRBinDiff}  & \textbf{0.960} & \textbf{0.915} & 0.897 & \textbf{0.909} & \textbf{0.847} & \textbf{0.861} & \textbf{0.838} & \textbf{0.890} &
    \textbf{0.992} & \textbf{0.990} & \textbf{0.992} & \textbf{0.987} & \textbf{0.986} & 0.983 & \textbf{0.980} & \textbf{0.987} \\ 
    w/o Norm & 0.958 & 0.898 & 0.872 & 0.894 & 0.779 & 0.826 & 0.787 & 0.859 
    & 0.988 & 0.968 & 0.971 & 0.970 & 0.947 & 0.952 & 0.953 & 0.964 \\
    w/o PLM & 0.939 & 0.880 & 0.843 & 0.866 & 0.746 & 0.770 & 0.739 & 0.826  
    & 0.984 & 0.965 & 0.954 & 0.959 & 0.929 & 0.938 & 0.925 & 0.951 \\
    w/o Graph & 0.928 & 0.839 & 0.527 & 0.846 & 0.476 & 0.514 & 0.506 & 0.662
    & 0.985 & 0.952 & 0.861 & 0.961 & 0.842 & 0.853 & 0.876 & 0.904 \\
    \hline \bottomrule
    \end{tabular}
        \begin{tablenotes}
            \item[1] Asm2Vec and PalmTree can not support cross-architecture (XA) tasks.
        \end{tablenotes}
\end{threeparttable}
}
\label{tab:recall1050}
\end{table*}

The Recall@10 and Recall@50 in One-to-many search on unbalanced set are shown in Table \ref{tab:recall1050}.

\end{document}